\documentclass{article}
\usepackage{graphicx, subfigure}
\usepackage{amsmath}
\usepackage{enumerate}

\title{Periastron precession for an extremal spherically symmetric dilaton black hole}

\author{Paul Blaga\\Babe\c{s}-Bolyai University\\Faculty of Mathematics and Computer Science\\ Kog\u{a}lniceanu Street 1, 600410, Cluj-Napoca, Romania\\
	Email: pablaga@cs.ubbcluj.ro\\
	Cristina Blaga\\Babe\c{s}-Bolyai University\\Faculty of Mathematics and Computer Science\\
	Kog\u{a}lniceanu Street 1, 600410, Cluj-Napoca, Romania\\
	Email: cpblaga@math.ubbcluj.ro}

\begin{document}
\maketitle

\begin{abstract}
The purpose of this article is to obtain the periastron precession of a free particle  moving around an extremal spherically symmetric dilaton black hole. To get the formulae for the periastron precession we use the phase-plane analysis of the general relativistic equations of motion.	
\end{abstract}

\section{Introduction}

The metric of the spherically symmetric dilaton black hole was found by Gibbons and Maeda \cite{gm} and, independently, by Garfinkle, Horowitz and Strominger \cite{ghs}.  Therefore, this solution of the Einstein-dilaton field equations is known as the Gibbons-Maeda-Garfinkle-Horowitz-Strominger (GMGHS) black hole.

The corresponding line element is
\begin{equation}\label{metr}
ds^2=-\left(1-\frac{2M}{r}\right) dt^2 +\frac{d r^2}{\left(1-\dfrac{2M}{r} \right)} + r \left(r-\frac{Q^2}{M}\right) \left(d \theta^2 + \sin^2 \theta \, d \varphi^2\right),
\end{equation} 
where $Q$ is a parameter related to the electrical charge of the black hole, while $M$ to its mass. The parameters $M$ and $Q$ are such that $Q^2\leq 2M^2$, otherwise the spacetime would not describe a black hole (there wouldn't be an event horizon). The case $Q^2=2M^2$ is known as an \emph{extremal black hole}.\footnote{We use here the geometric units, in which $c=G=1$, therefore $r_S=2M$, where $r_S$ is the Schwarzschild radius.}

We will focus in this paper on the case of extremal dilaton black hole. To be more specific, we consider here a \emph{massless} dilaton spherically symmetric black hole. For a discussion of the analogous spherically symetric solution with a \emph{massive} dilaton, see~\cite{hor}.

The GMGHS metric is a solution of the Einstein-Maxwell (massless) dilaton field equations. It depends on two parameters: the mass and the electric charge. There is another family of static, spherically symetric metrics, that also depend on the same parameters, the Reissner-Nordstr\"om metrics (see, for instance,~\cite{cha}), described by
\begin{equation}\label{rnmetr}
ds^2=-\left(1-\frac{2M}{r}+\frac{Q^2}{r^2}\right) dt^2 +\frac{d r^2}{\left(1-\dfrac{2M}{r}+\dfrac{Q^2}{r^2} \right)} + r^2  \left(d \theta^2 + \sin^2 \theta \, d \varphi^2\right).
\end{equation}
A Reissner-Nordstr\"om metric is a solution of the Einstein-Maxwell equation which, also, describes a spherically symmetric spacetime of an electrically charged black hole, provided $Q^2\leq M^2$. It also have an extremal case, corresponding to $Q^2=M^2$.

The two metrics correspond to different gravitation theories, but the are often compared (see~\cite{hor}). In the extremal case, they are considerable different. 

An extremal Reissner-Nordstr\"om black hole is still a black hole (in the sense that it still has a non-degenerate event horizon, hiding the singularity).

An extremal dilaton black hole, instead, is a more ``strange'' object. First of all, it is not, really, a black hole. Indeed, an easy computation shows that the area of the event horizon is equal to zero, \emph{i.e.} there is no regular event horizon. On the other hand, the $r-t$ plane of the spacetime is independent on $Q$ (in fact, it is identical to the Schwarzschild one). Therefore, the singularity is null, instead of being timelike, as it usually happens in the case of naked singularities. For a detailed discussion of these issues (also in terms of entropy and temperature) see~\cite{ghs}, \cite{hol} or \cite{hor}.
\section{The equations of motion}

The Lagrangean $\mathcal{L}$ corresponding to the line element~(\ref{metr}), given by
\begin{equation}\label{lagr}
2\mathcal{L}=-\left(1-\frac{2M}{r}\right)\dot{t}^2+\frac{\dot{r}^2}{1-\dfrac{2M}{r}}+r \left(r-\frac{Q^2}{M}\right) \left(\dot{\theta}^2 + \sin^2 \theta \dot{ \varphi}^2\right),
\end{equation}
is a constant of motion.
Here a dot means differentiation with respect to the affine parameter $\tau$ along the geodesic, chosen in such a way that $2\mathcal{L}=-1$ along timelike geodesics, $2\mathcal{L}=0$ along null geodesics and $2\mathcal{L}=1$ along spacelike geodesics.

The motion of the free test particles around a black hole takes place on timelike geodesics, therefore, hereafter, we shall assume that $2\mathcal{L}=-1$.

We derive the equations of motions of a free test particle by using the Euler-Lagrange equations associated to the Lagrangean~(\ref{lagr}) (see~\cite{cha}). Using this formalism, the equations of motion around a GMGHS black hole were obtained by several authors (\cite{bb}, \cite{f12}, \cite{ov}, \cite{b15}), emphasing that if in the beginning the free test particle lies in the equatorial plane $\theta=\pi/2$ and $\dot{ \theta}=0$, then the motion is confined to the equatorial plane, being, thus, planar. There are two additional constants of motion introduced by the integral of energy:
\begin{equation}\label{energ}
\left(1-\frac{2M}{r}\right)\dot{t}=E
\end{equation}
and the integral of angular momentum:
\begin{equation}\label{momen}
r\left(r-\frac{Q^2}{M}\right)\dot{ \varphi}=L,
\end{equation}
where $E$ is the total energy of the particle, while $L$ is its angular momentum about an axis normal to the plane of motion.

Using the constancy of the Lagrangean, we obtain that
\begin{equation}
\left(\frac{dr}{d\tau}\right)^2+\left(1-\frac{2M}{r}\right)\left(\frac{L^2}{r\left(r-\dfrac{Q^2}{M}\right)}+1\right)=E^2 \label{eq5}
\end{equation}
on a timelike geodesic. The second term from the left-hand side of the relation~(\ref{eq5}) is named the \emph{effective potential}, by analogy to the Newtonian mechanics.

The effective potential of an extremal GMGHS black hole is
\begin{equation}\label{eq6}
V_{eff}(r)=\frac{L^2}{r^2}-\frac{2M}{r}+1.
\end{equation}
From the equation~(\ref{eq5}), we observe that if $E^2\geq V_{eff}$, then the motion is possible. If $E^2=V_{eff}$, then $\dfrac{dr}{d\tau}=0$ and $r=\text{constant}$, meaning that the particle describes a circle in the equatorial plane. The circular orbits near a GMGHS black hole were studied in \cite{b13} or \cite{p15}. 

The effective potential~(\ref{eq6}) has some interesting properties: it tends to 1 when $r$ goes to infinity and it has a single extremal point, $r_e=L^2/M$.

Further we study the motion in the equatorial plane and seek the dependence of $r$ on the angle $\varphi$, therefore we transform the equation~(\ref{eq5}) using the integral of motion~(\ref{momen}). We get
\begin{equation}\label{eq8}
\left(\frac{dr}{d\varphi}\right)^2=\frac{r^2\left(r-r_S\right)^2}{L^2}\left(E^2-V_{eff}(r)\right),
\end{equation}
and using the notation $x=\dfrac{r_S}{r}$, we get, farther,
\begin{equation}\label{eq9}
\left(\frac{dx}{d\varphi}\right)^2=(1-x)^2\left[2\sigma\left(E^2-1\right)+2\sigma x - x^2\right],
\end{equation}
where $\sigma \equiv \dfrac{1}{2}\left(\dfrac{r_S}{L}\right)^2$ is a dimensionless parameter. A similar equation for the motion of a free test particle around a GMGHS black hole has been obtained by other authors\footnote{We get the equation~(\ref{eq9}) if we replace $Q^2=2M^2$ in the equation (10) from~\cite{bb18}.}.

\section{Phase plane analysis}
By differentiating~(\ref{eq9}) with respect to $\varphi$, we get the second order differential equation
\begin{equation}\label{eq10a}
\frac{d^2x}{d\varphi^2}=(1-x)\left[2x^2-(3\sigma+1)x-2\sigma E^2+3\sigma\right].
\end{equation}
To perform the phase-plane analysis, we write the equation~(\ref{eq10a}) as a system of two first order differential equations 
\begin{equation}\label{eq11}
\begin{cases}
\dfrac{dx}{d\varphi}=y,\\
\dfrac{dy}{d\varphi}=(1-x)\left[2x^2-(3\sigma+1)x-2\sigma E^2+3\sigma\right].
\end{cases}
\end{equation}
The equilibrium points of the system~(\ref{eq11}) are obtained by solving the system $x'=y'=0$ for $x$ and $y$. If $x'=0$, we get $y=0$. When $y'=0$, we get 
\begin{equation}\label{yp0}
(1-x)[2x^2-(3\sigma+1)x-2\sigma E^2+3\sigma]=0.
\end{equation}
But if $x'=0$, from~(\ref{eq9}), we obtain
\begin{equation}\label{yp08}
(1-x)^2[2\sigma\left(E^2-1\right)+2 \sigma x -x^2]=0.
\end{equation}
If $x \neq 1$, from equation~(\ref{yp08}) we get
\begin{equation}\label{E2}
2 \sigma E^2 = 2 \sigma - 2 \sigma x + x^2.
\end{equation} 
Replacing~(\ref{E2}) in the second equation of the system~(\ref{eq11}), the right hand side becomes $(1-x)^2 (\sigma-x)$. And so, we conclude that the system~(\ref{eq11}) has two equilibrium points: $(1,0)$ and $(\sigma,0)$.

Let us recall that $x=\dfrac{r_S}{r}$. It follows that the point with $x=1$ is on the event horizon and $x=\sigma$ corresponds to a point lying outside the event horizon if and only if $0<\sigma<1$. For $x=\sigma$, from the relation~(\ref{eq9}), we obtain that $E^2=1-\sigma/2$, which is equal to the minimum value of the effective potential. Therefore, the equilibrium point $(\sigma,0)$ of the system is stable.  

\section{Linear stability}

In order to obtain the nature of the equilibrium points of the system~(\ref{eq11}), we consider a small perturbation $\delta x=x-x^*$, $\delta y=y-y^*$, about a fixed point $\left(x^*,y^*\right)$  of the system 
\begin{equation}\label{eq12}
\begin{cases}
x'=y,\\
y'=(1-x)^2(\sigma-x).
\end{cases}
\end{equation}
The displacement from the equilibrium point is small, therefore we drop the second and higher order terms in $\delta x$, $\delta y$, $y^*=0$ and so we get the first order linear equations near the equilibrium point $\left(x^*,y^*\right)$:
\begin{equation*}
\begin{cases}
\delta x'=\delta y,\\
\delta y'=-\left(1-x^*\right)\left(-3x^*+2\sigma+1\right)\delta x.
\end{cases}
\end{equation*}
We write this system in matrix form
\begin{equation}\label{eq13}
\begin{pmatrix}
\delta x'\\
\delta y'
\end{pmatrix}
=
\begin{pmatrix}
0&1\\
-\left(1-x^*\right)\left(-3x^*+2\sigma+1\right)&0
\end{pmatrix}
\begin{pmatrix}
\delta x\\
\delta y
\end{pmatrix}
\end{equation}
where
\begin{equation}\label{eq14}
A\big|_{(x^*,0)}\big.=
\begin{pmatrix}
0&1\\
-\left(1-x^*\right)\left(-3x^*+2\sigma+1\right)&0
\end{pmatrix}
\end{equation}
is the matrix of the system. The nature of the equilibrium points depends on its eigenvalues. The trace of the matrix is $\mathcal{T}=0$ for any value of $x^*$, its determinant (depending on $x^*$) is
\begin{equation*}
\Delta =\left(x^*-1\right)\left(-3x^*+2\sigma+1\right),
\end{equation*}
while its eigenvalues are
\begin{equation*}
\lambda_{1,2}=\frac{1}{2}\left(\mathcal{T}\pm \sqrt{\mathcal{T}-4\Delta}\right)=\pm\sqrt{-\Delta}.
\end{equation*}

We notice that for the equilibrium point $(1,0)$ the matrix of the linear system becomes
\begin{equation*}
A\big|_{(1,0)}\big.=
\begin{pmatrix}
0&1\\
0&0
\end{pmatrix},
\end{equation*}
hence this case is a degenerate one.

For the second equilibrium point, $(\sigma,0)$, the matrix
\begin{equation*}
A\big|_{(\sigma,0)}\big.=
\begin{pmatrix}
0&1\\
-(1-\sigma)^2&0
\end{pmatrix}
\end{equation*}
has the complex eigenvalues
\begin{equation*}
\lambda_{1,2}=\pm \sqrt{-(1-\sigma)^2}=\pm i|1-\sigma|,
\end{equation*}
meaning that the equilibrium point is a center. The trajectories around a center correspond to elliptical orbits and for them we will compute the precession of the periastron. The system~(\ref{eq11}) is reversible, meaning that it is invariant at the transformation $t \rightarrow -t$ and $y \rightarrow -y$. The centers are robust for reversible systems, meaning that sufficiently close to the linear center $(\sigma,0)$ all trajectories are closed orbits (see \cite{s94}, page 164). 

\subsection{The precession of the periastron}
The trajectory around a center could be an ellipse. If the apsidal line rotates in the plane of motion, the ellipse described by the particle doesn't close after a full period. The motion of the apsidal line, around the center $(\sigma,0)$, can be investigated by using the linear approximation of the system~(\ref{eq12}), written like
\begin{equation}\label{eq15}
\begin{cases}
\delta'x=\delta y,\\
\delta'y=-\omega^2\delta x,
\end{cases}
\quad \text{where}\;\; \omega=|1-\sigma|.
\end{equation}
The solution of the system~(\ref{eq15}) corresponding to a precessing ellipse is
\begin{equation}\label{eq16}
\begin{cases}
\delta x(\varphi)=a\cos \omega\varphi+b\sin \omega\varphi,\\
\delta y(\varphi)=-\omega a\sin\varphi+\omega b\cos\omega\varphi ,
\end{cases}
\end{equation}
where $a$ and $b$ are integration constants, deduced from the initial data.

If we let $\delta x(0)=x_0$ and $\delta y(0)\equiv\dfrac{dx}{d\varphi}\Big|_0=0$, then we get, immediately, that $a=x_0$ and $b=0$, therefore the system~(\ref{eq16}) becomes
\begin{equation}\label{eq17}
\begin{cases}
\delta x=x_0\cos \omega \varphi,\\
\delta y(\varphi)=-\omega x_0\sin \omega \varphi.
\end{cases}
\end{equation}
The period of motion on the orbit~(\ref{eq17}), denoted by $\Phi$, is given by
\begin{equation}
\Phi=\frac{2\pi}{\omega}=\frac{2\pi}{|1-\sigma|}.
\end{equation}
If $\sigma$ is small, then 
\begin{equation}\label{pp}
\Phi \approx 2 \pi + 2 \pi \sigma,
\end{equation}
meaning that the apsidal line precess the motion of the particle. The periastron precession of a particle in a Schwarzschild field was determined by Dean~\cite{db99}. He obtained $\Delta \Phi_{Sch}=3 \pi \sigma$. Comparing his result with the relation~(\ref{pp}), we get that $\Delta \Phi_{dilaton}=-\pi \sigma$, meaning that the dilaton decelerates the periastron preccesion. Let us emphasize that Olivares and Villanueva \cite{ov} arrived to a similar conclusion using the method of derivation of perihelion advance proposed by Cornbleet~\cite{cb93}.

\section{Exact phase-plane}

The linear stability analysis gave us a clue about the motion in the neighborhood of the equilibrium points. If we want to obtain the features of the motion in the entire phase plane, we have to go back to the equation of motion~(\ref{eq9}). The left hand side of this equation is always greater or equal with zero. Thus, the motion is possible outside the events horizon ($x \geq1$) if $2\sigma\left(E^2-1\right)+2\sigma x - x^2 \geq 0$. If $dx/d\varphi=0$, the particle describes a circle, and $x=1$ or $2\sigma\left(E^2-1\right)+2\sigma x - x^2 = 0$. If $x=1$, the orbit is a circle with radius $r=r_S$, where $r_S$ is represent the events horizon.

The square root from the equation~(\ref{eq9}) leads us to 
\begin{equation}\label{eq9r}
\frac{dx}{d\varphi}=\pm |1-x| \sqrt{2\sigma\left(E^2-1\right)+2\sigma x - x^2},
\end{equation} 
an ordinary differential equation with separable variables. The $\pm$ sign form the right hand side of the equation~(\ref{eq9r}) is related to the sense in which the orbit is described.

Introducing the variable $z=1-x$, the equation~(\ref{eq9r}) becomes 
\begin{equation}\label{vsq}
\frac{dz}{|z| \sqrt{2\sigma E^2-1 + 2\left(\sigma -1\right)z - z^2}}=\mp d\varphi,
\end{equation} 
a differential equation with the general solution
\begin{equation}\label{eq24}
\mp (\varphi-\varphi_0)=
\begin{cases}
-\dfrac{1}{\sqrt{a}} \, \ln \dfrac{2\sqrt{aZ}+bz+2a}{z} \,\,,\quad a>0\\
-\dfrac{2 \sqrt{Z}}{bz} \,\,, \qquad a=0\\
\dfrac{1}{\sqrt{-a}} \, \arcsin \dfrac{bz+2a}{|z|\sqrt{-q}}\,\,, \quad a<0
\end{cases}
\end{equation}
where $Z=a+bz+cz^2$, $q=4ac-b^2$, $a=2 \sigma E^2-1$, $b=2(\sigma-1)$ and $c=-1$ \cite{b91}. We note that $q=-8 \sigma (E^2-1+\sigma/2)$ is always less or equal with zero, because the minimal value of the effective potential~(\ref{eq6}) is $1-\sigma/2$ and the motion is possible if and only if $E^2 \geq V_{eff}$. 

The closed orbits in the phase plane correspond to the third branch of~(\ref{eq24}). A particle describes such orbit, if its energy satisfies the condition $E^2 <1/(2 \sigma)$. The period of motion is $2 \pi/\sqrt{1-2 \sigma E^2}$. In the equilibrium point $(\sigma, 0)$, the energy is $E^2=1-\sigma/2$ and the corresponding period of motion becomes 
\begin{equation}
\Phi = \dfrac{2 \pi}{|1-\sigma|},
\end{equation} 
expression obtained using the linear stability analysis.    

\begin{figure}[ht!]
	\begin{center}
		\subfigure[$\sigma=1/5$]{%
			\label{fig:first}
			\includegraphics[width=0.45\textwidth]{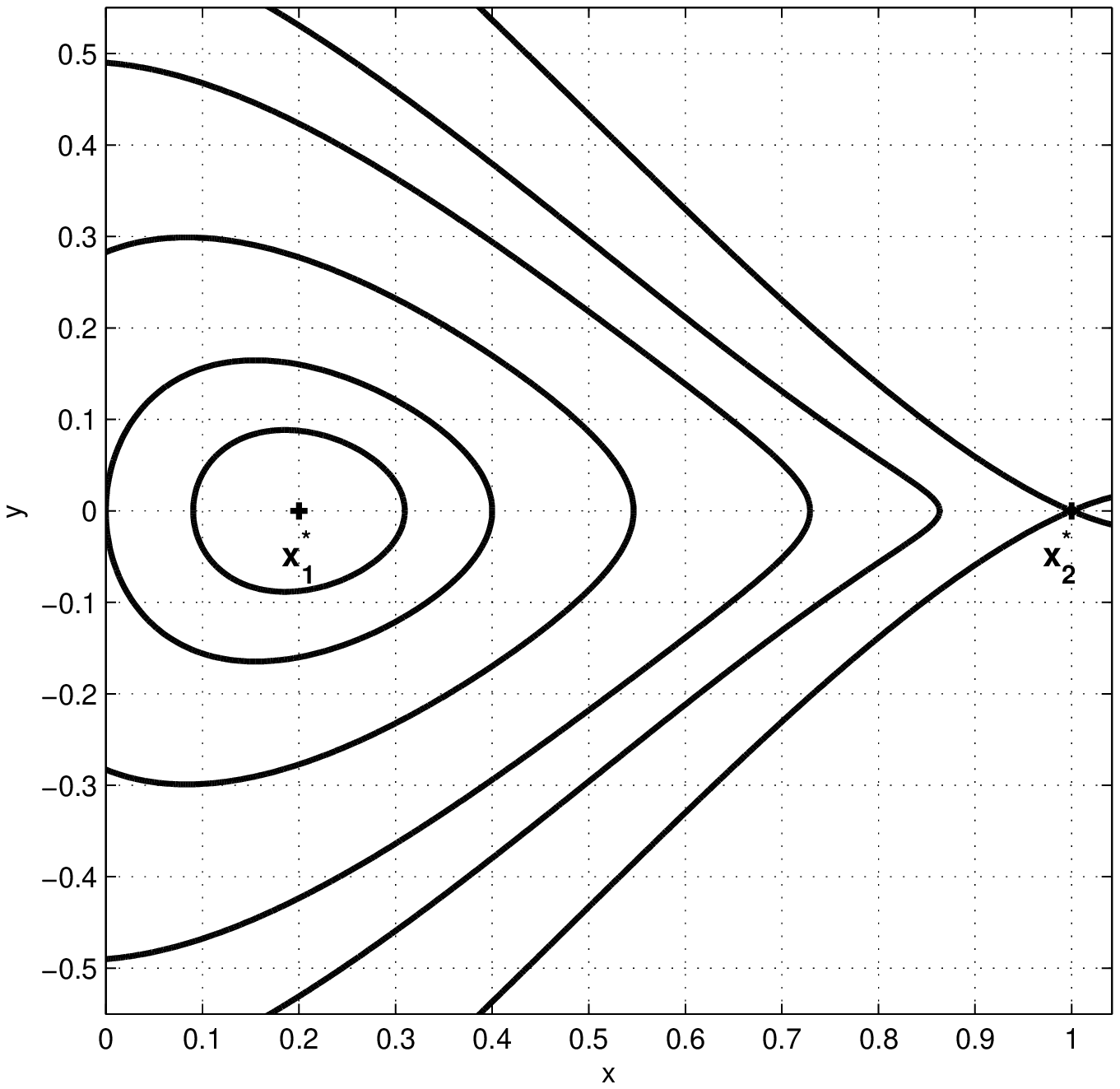}
		}%
		\subfigure[$\sigma=1/2$]{%
			\label{fig:second}
			\includegraphics[width=0.45\textwidth]{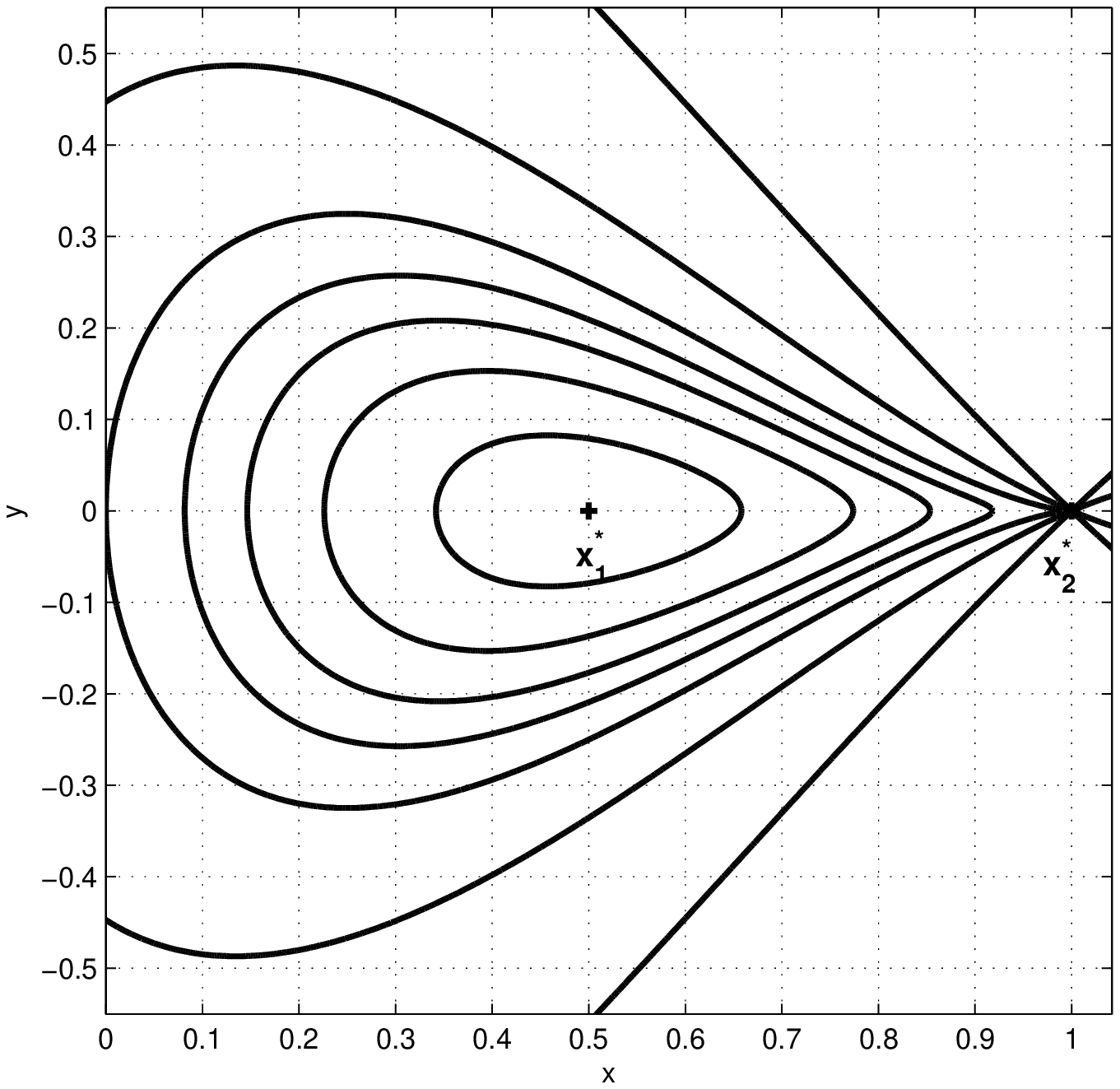}
		} %
	\end{center}
	\caption{%
		Phase portrait for different values of $\sigma$. 
	}
	\label{f1}
\end{figure}

In figure~{\ref{f1}} we have represented the phase plane diagram of the system~(\ref{eq11}) for different values of the parameter $\sigma$. The equilibrium points $x_1^{*}(\sigma,0)$,  $x_2^{*}(1,0)$ are represented with $+$ and with solid lines the orbits for different values of the energy. The first equilibrium point moves to the events horizon when $\sigma \rightarrow 1$.



%

\end{document}